\documentclass[twocolumn,
groupedaddress,
superscriptaddress,
prl,
]{revtex4-1}

\usepackage{amsmath}
\usepackage{subfigure}
\usepackage{mdframed}
\usepackage{graphicx}
\usepackage{hyperref}
\hypersetup{
	colorlinks = true, linkcolor = blue, anchorcolor = blue, urlcolor = blue, citecolor = blue
}

\usepackage{tablefootnote}

\begin{document}
	
	
    \title{Provably-Secure and High-Rate Quantum Key Distribution with Time-Bin Qudits}
	
	
	\author{Nurul T. Islam}
	\affiliation{Department of Physics and the Fitzpatrick Institute for Photonics, Duke University, Durham, North Carolina 27708, USA}
	\email{nti3@duke.edu}
	\author{Charles Ci Wen Lim}
	\affiliation{Computational Sciences and Engineering Division,
	Oak Ridge National Laboratory, Oak Ridge, TN 37831-6418, USA} \email{charles.lim@nus.edu.sg}
	\affiliation{Department of Electrical and Computer Engineering,
National University of Singapore, 4 Engineering Drive 3, Singapore 117583}
	\author{Clinton Cahall}
	\affiliation{Department of Electrical Engineering and the Fitzpatrick Institute for Photonics, Duke University, Durham, North Carolina 27708, USA}
	\author{Jungsang Kim}
	\affiliation{Department of Electrical Engineering and the Fitzpatrick Institute for Photonics, Duke University, Durham, North Carolina 27708, USA}
	\author{Daniel J. Gauthier}
	\affiliation{Department of Physics, The Ohio State University, 191 West Woodruff Ave., Columbus, Ohio 43210 USA}



\begin{abstract}
 The security of conventional cryptography systems is threatened in the forthcoming era of quantum computers. Quantum key distribution (QKD) features fundamentally proven security and offers a promising option for quantum-proof cryptography solution. Although prototype QKD systems over optical fiber have been demonstrated over the years, the key generation rates remain several orders-of-magnitude lower than current classical communication systems. In an effort towards a commercially viable QKD system with improved key generation rates, we developed a discrete-variable QKD system based on time-bin quantum photonic states that is capable of generating provably-secure cryptographic keys at megabit-per-second (Mbps) rates over metropolitan distances. We use high-dimensional quantum states that transmit more than one secret bit per received photon, alleviating detector saturation effects in the superconducting nanowire single photon detectors (SNSPDs) employed in our system that feature very high detection efficiency (of over 70\%) and low timing jitter (of less than 40 ps).  Our system is constructed using commercial off-the-shelf components, and the adopted protocol can readily be extended to free-space quantum channels.  The security analysis adopted to distill the keys ensures that the demonstrated protocol is robust against coherent attacks, finite-size effects, and a broad class of experimental imperfections identified in our system.
\end{abstract}


\maketitle

Development of scalable quantum computing platforms is one of the rapidly expanding areas of research in quantum information science \cite{Monroe2016,Monroe2017}. With many commercial companies working towards building these platforms, a medium-scale quantum computer capable of demonstrating quantum supremacy over classical computers is in earnest only a few years away. Quantum computers poses a serious threat to the cybersecurity because most of the current cryptosystems, like the one devised by Rivest, Shamir and Adleman (known as the RSA)--whose security is based on computational hardness assumptions---can potentially be broken with a powerful quantum computer in practical timescales \cite{Chuang05,Ahsan15}. Quantum key distribution (QKD) with symmetric encryption is one of the very few methods that can provide provable security against an attack aided with a quantum computer \cite{BarnettBook}. However, a major limitation of most current QKD systems is that the rate at which the secret key is generated is orders-of-magnitude lower than the digital communication rates~\cite{Lo14}.~This limitation ultimately prevents QKD from being useful for a wide range of communication tasks.

To make QKD more relevant for widespread deployment in communication networks, there has been significant effort to increase the key generation rate of QKD systems, prioritizing metropolitan distances (20-80~km) for large-scale implementation of QKD networks \cite{Diamanti16}. One of the major breakthroughs was the development of superconducting nano-wire single-photon detectors that can detect photons with high-efficiency and yet have low dark count rates \cite{Marsili13}. However, these detectors still have a recovery time greater than 10~ns \cite{Zhao14}, thereby limiting the rate at which the secret key can be generated. 

High-dimensional quantum states---qudits (dimension $d>2$) rather than qubits---provide a robust and efficient platform to overcome some of the practical challenges of current QKD systems~\cite{Bechmann01,Cerf02}. The efficiency comes from the ability to encode many bits ($\log_2d$) of information on a single photon. QKD systems using a high-dimensional quantum state space relies on the same degrees-of-freedom as the qubit-based systems. Nonetheless, the amount of information that can be encoded on each photon can be large even in a realistic situation because the number of bits that can be encoded on each photon is unbounded, scaling as $\log_2(d)$.

Fundamentally, QKD systems using a high-dimensional quantum state space have two major advantages over the qubit-based protocols. First, they can increase the effective key generation rate in systems limited by the saturation of the single photon detectors, often arising from the “dead time” of the detectors. The dead time refers to the period of time over a which a single-photon detector resets from a prior detection event and thus remains unresponsive to an incident photon. This becomes particularly important in the limit of low channel loss, which corresponds to relatively short distances in standard optical fiber. Second, high-dimensional QKD systems have higher resistance to quantum channel noise, which means these systems can tolerate a higher quantum bit error rate compared to qubit-based systems \cite{Valero10}.
\begin{figure*}[ht]
	\begin{center}
		\includegraphics[width= 0.7\textwidth]{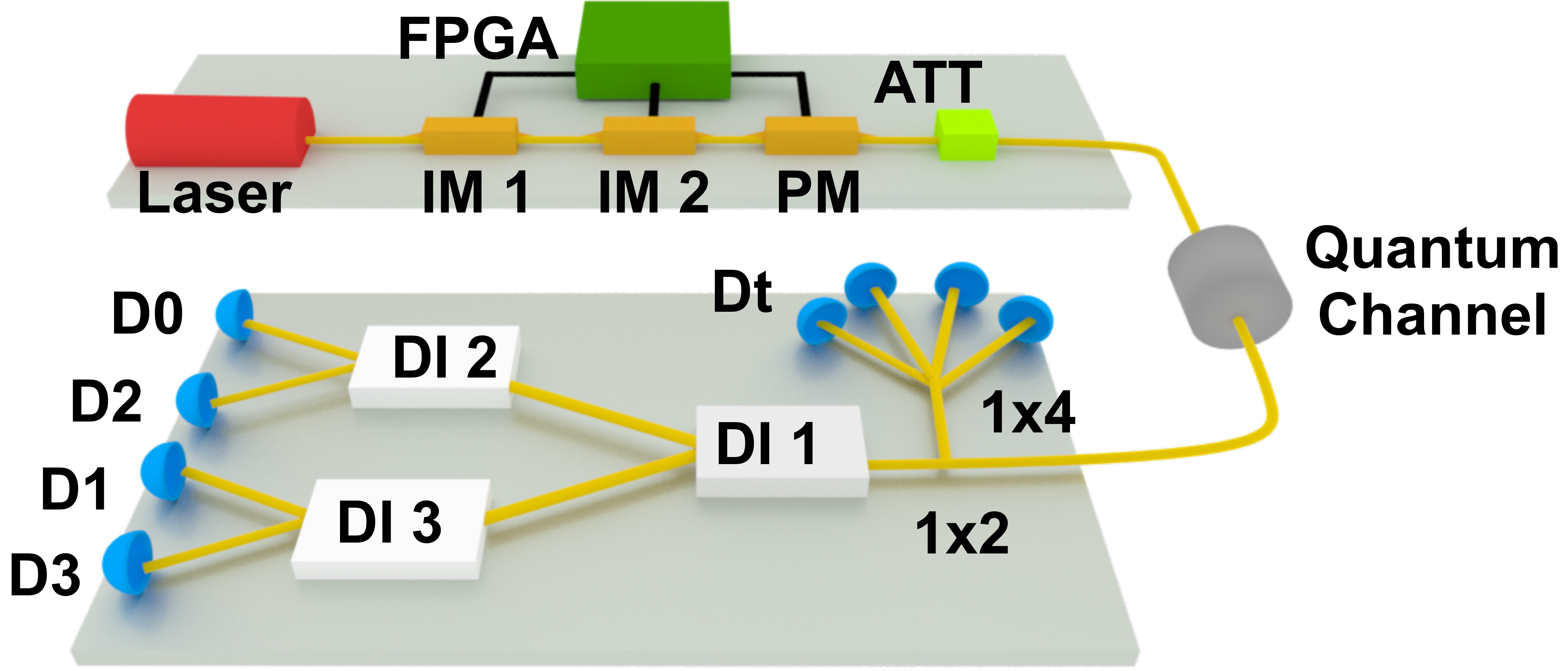}
		\caption{\textbf{Schematic of the experimental setup.} At Alice's transmitter, the quantum photonic states (signal and decoy) are created using a frequency-stabilized continuous laser (Wavelength Reference, Clarity-NLL-1550-HP) operating at 1550~nm, which passes through three intensity modulators (only two are shown for clarity) and one phase modulator (all intensity and phase modulators are from EOSpace). The entire system is controlled by serial pattern generators realized with a field-programmable gate array (FPGA, Altera Stratix V 5SGXEA7N2F40C2), operating at a 10 GHz clock rate. In greater detail, a 5~GHz sine-wave generator phase locked to the FPGA drives an intensity modulator (not shown), which creates a periodic train of 66-ps-duration (full width at half maximum) optical pulses. These pulses pass through an intensity modulator (IM 1), which is driven by the FPGA-based pattern generator to define the data pattern for either the time-bin or phase states. A second intensity modulator (IM 2), driven by an independent FPGA channel, adjust the amplitude of the phase and decoy states relative to the primary time-bin signal states.  Finally, the states pass through an FPGA-driven phase modulator (PM) to encode the different phase states. The time-bin basis and the phase basis are chosen with probabilities 0.90 and 0.10, respectively. An attenuator (ATT) reduces the level of the states to the single-photon level.  An additional attenuator is used to simulate the loss of the quantum channel.  At Bob's receiver, the incoming signals are split using a $90/10$ beamsplitter (BS) to direct 90\% of the states to the temporal basis measurement system and 10\% to the phase basis system. For both measurement bases, we use commercially available superconducting nanowire single-photon detectors (Quantum Opus) and the detection events are recorded with a 50-ps-resolution time-to-digital converter (Agilent, Acqiris U1051A), which is synchronized with Alice's clock over a public channel.}
		\label{fig:ExperimentalSetup}
	\end{center}
\end{figure*}

High-dimensional QKD systems have been demonstrated using various degrees-of-freedom of the photon, such as spatial~\cite{Leach12,Etcheverry13, Boyd14,HD16,Ding16} or time-energy modes~\cite{Brougham13,Shapiro13,Dan14,MIT15,Brougham15, MIT16}. Here, we use the photon's temporal degree-of-freedom because it is relatively unaffected by turbulence in a free-space channel and easily propagates through metropolitan-scale fiber networks. Using a four-dimensional ($d=4$) state space represented by four distinct time bins and it’s conjugate state space in the  Fourier transform domain, we realize a QKD that generates and ultra-high secret key rate. We note that our system is built using commercial off-the-shelf components, and therefore it can readily be realized using equipment found in many existing QKD systems. 

\section{Results}
Our QKD system is based on a prepare-and-measure scheme, where Alice randomly modulates a continuous-wave laser and attenuates the outgoing photonic wavepackets to the single-photon level.~The photonic wavepackets are then transmitted via an untrusted quantum channel to a distant receiver, called Bob, who uses single-photon detectors or interferometers coupled to single-photon detectors to measure the wavepackets in the time or phase bases, respectively. In addition, to deal with the so-called photon-number-splitting attacks, we use a practical decoy-state method to estimate the number of single-photon wavepackets received by Bob~\cite{Hwang03,Lodecoy05,Ma05,CharlesDecoy}. The secret key is calculated using the sifted photon time-of-arrival data, and the amount of extractable secret data is determined using the noise level observed in the sifted phase measurement data. An illustration of our experimental system is shown in Fig.~\ref{fig:ExperimentalSetup}.

\begin{figure*}[tbh]
	\begin{center}
		\includegraphics[width= 0.9\textwidth]{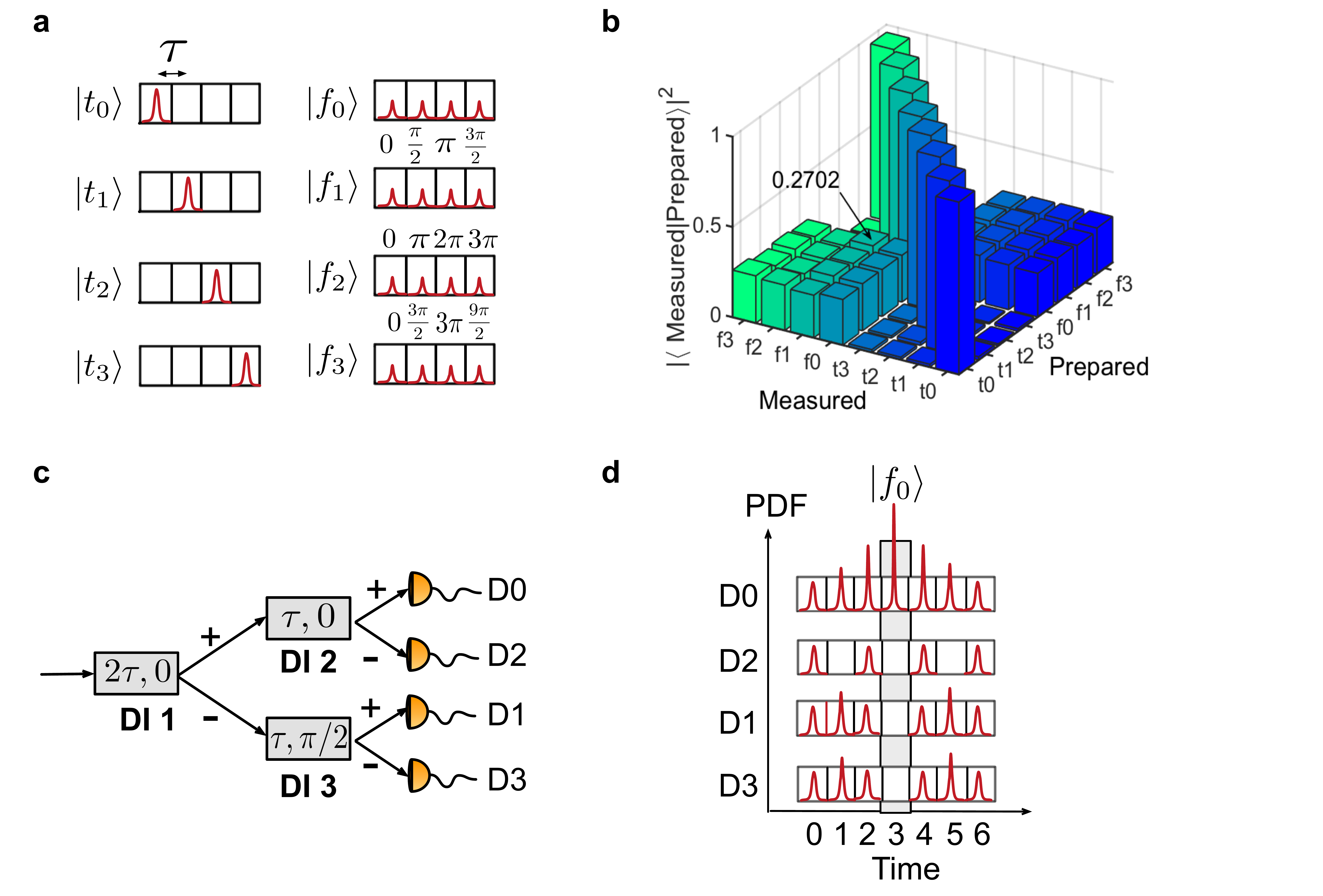}
		\caption{\textbf{Time-bin and phase states for $d = 4$ and the phase- state measurement scheme.} \textbf{a,} Temporal (left panel) and phase (right panel) states for $d = 4$ with the phases determined from Eq. \ref{DFT}. \textbf{b,} Probability of detection when each input state is measured in both bases. \textbf{c,} Measuring the phase states with a cascaded interferometric tree, where the relative time delay of the first unequal-path delay-line interferometer (DI~1) is twice the delay of DI 2 and DI 3. The phase of DI~3 is set to $\pi/2$. \textbf{d,} Expected photon probability distribution at the output of the interferometers when the phase state $|f_0\rangle$ is injected into the system. }
		\label{fig:States}
	\end{center}
\end{figure*}

The quantum eigenstates in a $d$-dimension time basis are denoted by $|t_n\rangle$ ($n = 0, ... , d-1$). Each eigenstate is represented by a photonic wavepacket of width $\Delta t = 66$~ps, well localized to a time bin $n$ of width $\tau = 400$~ps within a frame of $d$ contiguous time bins, as shown in Fig. \ref{fig:States}a for $d=4$. For fixed $\tau$, the maximum mutual information per received state between Alice and Bob scales as (log$_2 d)/d$ assuming there is no detector saturation. This quantity is identical for $d=2$ and $d=4$, but decreases for larger $d$.

When taking into account detector saturation in a high-rate system such as ours, the rate scales as (log$_2d$) assuming that the state (frame) duration $\tau d$ matches the characteristic detector saturation time (\textit{e.g.}, detector deadtime) and hence higher-dimension protocols outperform qubit ($d=2$) protocols~\cite{MIT16}. Furthermore, higher-dimension protocols have better noise tolerance, resulting in a higher secret key rate as discussed below. In our experimental implementation, we focus on $d=4$. 

To secure the QKD system, we use $d$-dimension phase states.  They are a linear superposition of all of the temporal states weighted by a unit-magnitude exponential phase factor given by
\begin{align}
|f_n\rangle = \frac{1}{\sqrt{d}} \sum\limits_{m= 0}^{d-1} \exp\left(\frac{2\pi i n m}{d}\right)|t_m\rangle~~~n = 0, ..., d-1 \label{DFT}
\end{align}
and illustrated in Fig.~\ref{fig:States}a. They take the form of the discrete Fourier transforms of the temporal states, have a multi-peaked spectrum with peak spacing $1/\tau$ and width $\sim 1/2\Delta t$ and the carrier frequency of each is shifted with respect to the others. The phase states are mutually unbiased with respect to the temporal states in that states prepared in one basis and measured in the other result in a uniformly uncertain outcome: $|\langle t_n| f_m\rangle|^2 = 1/d$. The bars along the anti-diagonal in Fig.~\ref{fig:States}b represent the experimentally determined values of these probabilities when a state is prepared and measured in different bases. 


At Bob's receiver, a beamsplitter is used to randomly direct the incoming quantum photonic wavepackets to either a temporal or phase measurement device. We measure the temporal states using high detection efficiency single-photon detectors with a temporal resolution better than 40~ps. The detector efficiency begins to drop when the detection rate exceeds 2~Mcounts/s due to the finite detector reset time (Sec. 3, Supplementary Information). To overcome this issue, we use a 1:4 coupler to randomly direct photons to one of four detectors, allowing us to operate at high rates occurring at lower channel loss.

A novel feature of our QKD system is the phase-state measurement device~\cite{Brougham13,Taimur2016} as shown in Fig.~\ref{fig:States}c. Each output of the interferometers is uniquely related to one of the phase states.  As illustrated in Fig.~\ref{fig:States}d, the relevant time bin for observing interference is the central time bin (time bin 3) and when a phase state $|f_n\rangle$ is incident in the interferometric setup, the central time bin emerging from detector D$n$, $n \in \{0,3\}$, experiences constructive interference from the superposition of all $d$ wavepackets and destructive interference in all other outputs. We use commercial delay interferometers that are designed to be field-deployable and hence require no active path-length stabilization.

The security of our QKD system is derived using a recently developed technique based on entropic uncertainty relations for qudits~\cite{Renner08, Tomamichel12}. Unlike previous analyses for high-dimensional QKD, our approach gives finite-key bounds for mutually unbiased states and is secure against general (coherent) attacks. To extract a secret key from the single-photon states, we use a three-intensity decoy-state method to estimate the single-photon statistics observed in the data. We thereby obtain a bound on the extractable secret key length in terms of the measured data as quantified by Eq.~\ref{SKL} in the Methods section.

\begin{figure}[h]
	\begin{center}
		\includegraphics[width = 0.5\textwidth]{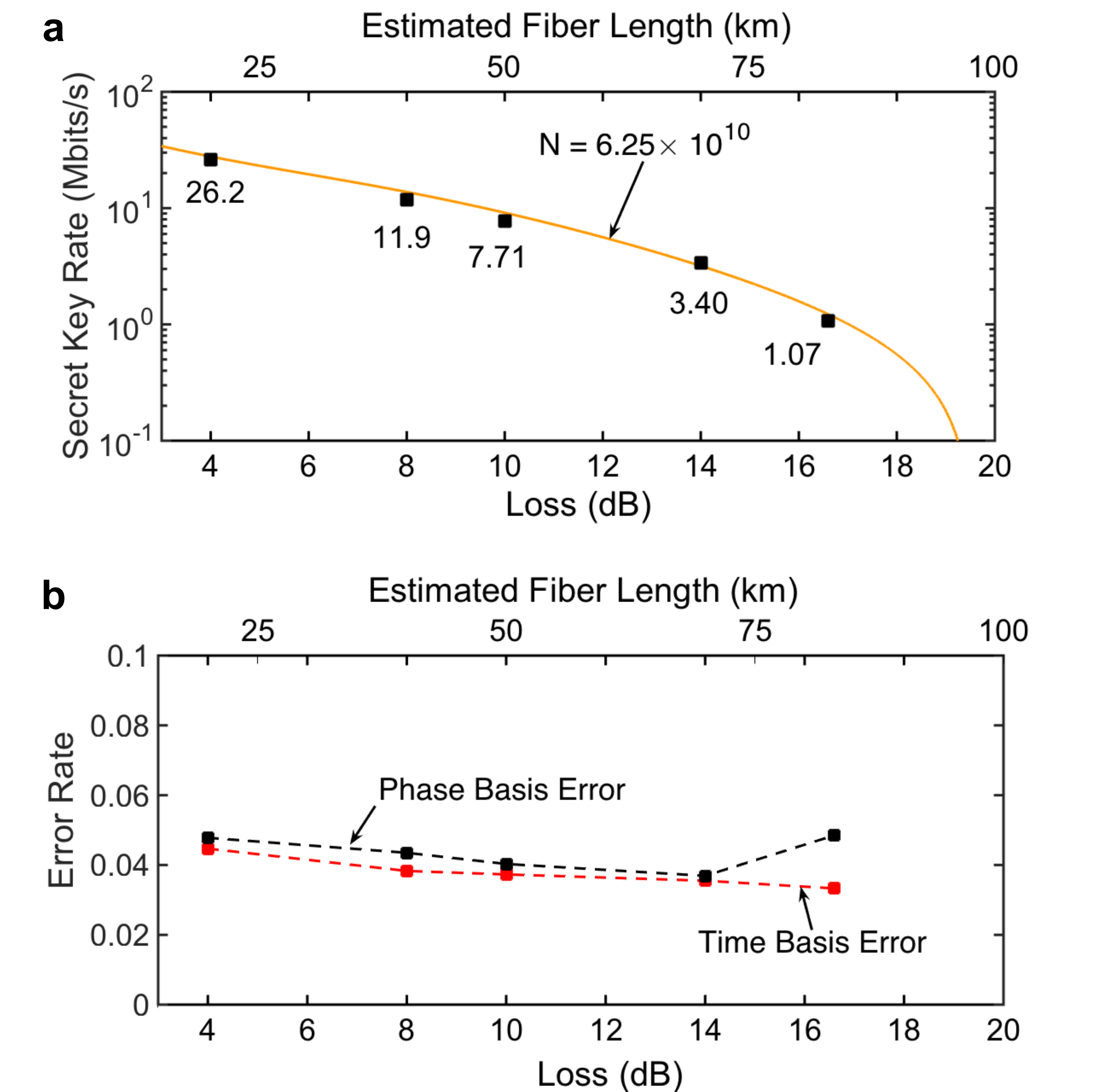}
		\caption{\textbf{Observation of high-rate secure quantum key distribution.} \textbf{a,} Experimentally achievable secret key rates as a function of the channel loss for the case when the number of signals transmitted by Alice is $N = 6.25\times 10^{10}$ (100-s-duration communication session). The orange solid line is the simulated secret key rate. For the simulation, we set the probabilities of sending signal, decoy and vacuum intensities to 0.8, 0.1, 0.1, respectively. The intrinsic error rate in the time and phase basis are set to 0.03 and 0.025, respectively. \textbf{b,} Experimentally observed quantum bit error rate in temporal and phase bases signal states as a function of channel loss.}
		\label{fig:Results}
	\end{center}
\end{figure}

Incorporating all of our experimental and theoretical tools, we realize a QKD system that is capable of generating record-high secret key rates. Our achieved secret key rate as a function of channel loss is shown in Fig.~\ref{fig:Results}a. For comparison to previous studies (Table~\ref{Performance}), we also represent the channel loss in terms of an equivalent length of optical fiber at telecommunication wavelengths (0.2 dB/km). At a channel loss of 4 dB (equivalent to a 20 km long optical fiber), we are able to achieve a secret key rate of 26.2 Mbits/s, which is the highest secret key rate reported at this quantum channel loss. For this case, the error rate in the temporal and frequency bases is 4.5\% and 4.8\%, respectively, as shown in Fig. \ref{fig:Results}b. We also obtain record-high secret key rates for other channel conditions up to to a loss of 16.6~dB (83 km) as illustrated in Table~\ref{Performance}.

The solid curve in Fig.~\ref{fig:Results}a is the simulated secret key rate obtained using experimentally observed parameters (see Methods). At the highest channel loss considered in the experiment (16.6 dB, 83 km), the detection rate is low enough that the detectors operate at their highest detection efficiency ($>70\%$). As the loss decreases, the detectors experience increasingly lower detection efficiency because of the finite detector reset time.  To account for this, we characterize the efficiency as a function of detection rate and incorporate this information in the security analysis (Sec. 3, Supplementary Information).

From the simulation, we see that the secret key rate drops rapidly beyond a loss of $18$ dB (90 km). This drop mainly occurs due to finite-key effects arising from our use of a fixed data collection interval for all data points.  In this case, the total data received by Bob goes down for higher channel loss, which increases the statistical uncertainty about the phase error rate (see Methods).

\section{Discussion}
We are able to obtain such high secret key rates due to multiple factors. First, for low-loss channels, the rate is ultimately limited by detector saturation. A high-dimensional protocol such as ours allows us to extract more bits per received photon at detector saturation in comparison to a qubit ($d=2$) protocol, essentially doubling the secret key rate for our $d=4$ protocol. Second, we use high-efficiency superconducting nanowire detectors that have a relatively short reset-time in comparison to other detectors operating in the telecommunication band, such as Geiger-mode avalanche photodiodes~\cite{Boris2014}.  Third, our detectors have nearly constant jitter ($<$ 40 ps) and low dark counts (100-200 counts/s) independent of detection rate, resulting in a nearly constant quantum bit error rate as a function of loss seen in Fig. \ref{fig:Results}b. Fourth, we match $\tau$  to be only somewhat larger than the detector jitter, allowing us to run at a high system clock rate of 2.5 GHz.

Our time-phase-state protocol is particularly well suited for field deployment because optical turbulence in free-space channels does not cause scattering of one of our photonic states into another as long as the wavepacket duration is substantially longer than 10 ps for path lengths of 10's of kilometers~\cite{Hamal05}.  Also, in a fiber-based system, the typical dephasing time is substantially longer than our frame duration time $\tau d$.

There are several possible directions for increasing the secret key rates in our system.  One is developing monolithic (possibly chip-based) interferometer trees to decrease the insertion loss (and hence decrease the phase error rate) and to increase $d$~\cite{MIT16, Brougham13}.  Another is to use dense-wavelength division multiplexing methods, where Alice uses multiple transmitters each with a different carrier frequency sent down the same quantum channel~\cite{Pan16}.  The  delay interferometers work across the entire telecommunication C-band and hence it should be possible to operate using multiple spectral channels with a single set of interferometers.  Such a system will require a large number of single-photon counting detectors, but substantial progress is underway in realizing arrays with 100's of detectors~\cite{Shaw15}.  Finally, there is considerable ongoing research in increasing the saturated detection rate of superconducting nanowire detectors~\cite{Zhao14}, which will have a major impact on any QKD system.

\begin{table*}[t]
	\caption{Comparison of some notable high-rate QKD systems.}
	\label{Performance}
	\begin{tabular}{|c|c|c|c|c|c|}
		\hline
		\hline
		&Protocol&Loss (dB)&Equivalent &Secret Key&~~Security Level~~\\
		& & &~~Fiber Length (km)~~&~~Rate (Mbits/s)~~& \\
		\hline
		~~Ref.\cite{Lucamarini13}~~ &T12&~~7~~&35&~~2.20~~&~~Collective\footnote[1]{For definitions of collective attack and coherent attacks, we refer readers to Ref.\cite{Valero09}.}~~\\
		& &~~10~~&50&~~1.09~~&\\     
		& &~~13~~&65&~~0.40~~&\\ 
		& &~~16~~&80&~~0.12~~&\\ 
		\hline
		~~Ref.\cite{MIT15}~~ &HD-QKD& 		~~0~~&0&~~7.0~~&~~Collective~~\\
		& &~~4~~&20&~~2.7~~&\\
		\hline
		~~Ref.\cite{MIT16}~~ &HD-QKD&			~~0~~&0&~~23.0~~&~~Collective~~\\
		& &~~8.2~~&41&~~5.3~~&\\
		& &~~12.7~~&~63\footnote[2]{43~km spool with 12.7 dB loss}&~~1.2~~&\\
		\hline
		~~Our work~~ &HD-QKD&~~4~~&20&~~26.2~~&~~Coherent/General~~\\
		& &~~8~~&40&~~11.9~~&\\
		& &~~10~~&50&~~7.71~~&\\
		& &~~14~~&70&~~3.40~~&\\
		& &~~16.6~~&83&~~1.07~~&\\
		\hline
		\hline
	\end{tabular}
\end{table*}

\section{Materials and Methods}
\noindent\textbf{Sketch of Security Proof}\\
The security of our QKD protocol is  defined by two criteria, namely the secrecy and correctness parameters, which we denote by $\epsilon_{\textsf{sec}}$ and $\epsilon_{\textsf{cor}}$, respectively. Using these criteria, we say that our protocol is $\varepsilon$-secure if it satisfies $\epsilon_{\textsf{sec}}+\epsilon_{\textsf{cor}}\leq\varepsilon$, where $\varepsilon$ is a predetermined security parameter. The correctness parameter $\epsilon_{\textsf{cor}}$ is typically fixed and determined by the length of hash codes used in the error verification step. Importantly, this choice of security definition guarantees that our QKD system is composable with any (possibly larger) cryptographic protocol, \textit{e.g.}, the one-time pad encryption protocol.  

Using these results, we find that the secret key length $\ell$ is given by
\begin{align}\label{SKL}
\ell \leq \max_{\beta\geq 0} \lfloor 2 \tilde{s}_{\mathsf{T},0} + \tilde{s}_{\mathsf{T},1}[c - H({\lambda^\textsf{U}})] - \textsf{leak}_{\textsf{EC}} + \Delta_\textsf{FK}\rfloor,
\end{align}
where $\tilde{s}_{\mathsf{T},0}$ and $\tilde{s}_{\mathsf{T},1}$ are the number of vacuum and single-photon detections in the raw key, respectively, and $\lambda^\textsf{U}$ is an upper-bound on the single-photon \emph{phase error rate} in terms of the observed error rate in the phase basis. The quality of the prepared states is quantified by the overlap parameter $c:=-\log_2 \max_{i,j}|\langle f_i|t_j \rangle|^2$.

During the calibration of our experiment, we measure a lower bound on this quantity of $c = 1.89$ as shown in Fig.~\ref{fig:States}d, where we plot the probability of detection matrices for all input states. Specifically, we measure all eight states in both basis and calculate the overlap of the prepared and measured states. When a state is measured in the same basis in which it was prepared, the probability should be $\sim$1, as indicated by the data along the diagonal. The quantity $c$ corresponds to the logarithm of the maximum of the anti-diagonal elements, where the measurement and preparation bases are different. For ideal state preparation and measurement, the overlap is 1/4, corresponding to $c = 2$; however, in the experiment, these matrix elements vary about 1/4 and we pick the element that gives the worst case estimate of $c$, as required by the overlap parameter defined above.

Finally, $H(x):= -x \log_2\left(x/3\right) - (1-x)\log_2(1-x)$ is the Shannon entropy for $d = 4$, $\textsf{leak}_{\textsf{EC}}$ is the number of bits published during error correction, and $\Delta_\textsf{FK}:= -\log_2 \left(32\beta^{-8} \epsilon^{-1}_{\textsf{cor}}\right)$.~The secret key length is maximized numerically over $\beta$ satisfying $4\epsilon_{\textsf{cor}} + 18\beta \leq \varepsilon$ (Sec. 1, Supplementary Information). 

\noindent{\textbf{Phase States Detection}}\\
We describe here our method for measuring the phase states because this system has not yet been widely discussed in the literature. The interferometric setup required to perform the frequency measurement consists of a cascade of three interferometers as shown in Fig.\ref{fig:States}c, where the second stage of the tree has interferometers whose time-delay ($\tau$) is a factor of two shorter than the interferometer in the first stage ($2\tau$)~\cite{Taimur2016}. 

When a phase state $|f_n\rangle$ ($n = 0, 1, 2, 3$) enters the interferometric setup, the first 50/50 beam splitter (BS) of DI 1 splits the wavepacket into two equal parts, with one part propagating through a longer arm relative to the other. The longer arm of the interferometer is set to delay the propagation of the wavepacket by $2\tau$ (two time bins) relative to the part propagating through the shorter arm. The two parts of the wavepacket then recombine at a second 50/50 BS in DI 1, resulting in an interference pattern at the two outputs denoted by $+$ and $-$. 

In the second stage interferometers, the wavepackets propagating through the longer arms are delayed by just one time bin before interfering with the part propagating through the shorter arms. The expected interference patterns, representing the probability distribution function (PDF) of the single-photon wavepackets, when state $|f_0\rangle $ propagates through the interferometric setup is shown in Fig.~\ref{fig:States}d.

It is seen that the wavepackets emerging from the interferometers occupy 7 time bins, where there is a 75\% chance that a photon is detected outside the central time bin in each channel. The central time bin is due to the interference of all four wavepacket peaks of the incident state and there is a one-to-one correspondence between the incident phase states, $|f_n\rangle$ and detection events in this time bin for detector D$n$.  We only use these events in our security analysis.  Except for the outermost peaks, the other peaks are due to interference of a sub-set of the incident wavepacket peaks. Although some information about the incident state can be extracted from measurement of photons in these peaks, we do not consider this here. 

A detailed analysis reveals that the sifting process ensures that there is no increase in error rate due to the spill-over of these wavepackets into the neighboring frames. However, the lower probability for a detection event in the central time bin reduces the overall number of events used in our security analysis and hence lowers our secret key rate.  On the other hand, the higher-dimension protocol used here has higher noise tolerance and allows for a higher secure rate~\cite{Valero09}.
\\
\noindent{\textbf{Acknowledgments}}
We gratefully acknowledge the financial support of the ONR MURI program on Wavelength-Agile Quantum Key Distribution in a Marine Environment, Grant \# N00014-13-1-0627, and the DARPA DSO InPho program. C.C.W.L acknowledges support from the Oak Ridge National Laboratory, operated by UT-Battelle for the U.S. Department of Energy under Contract No. DE- AC05-00OR22725 and the support
from NUS start-up grant R-263-000-C78-133/731. We acknowledge discussion of this work with Paul Kwiat, Andres Aragoneses and thank Daniel Kumor for providing us custom time-tagger data-collection software. \\

\bibliography{sample.bib}


\end{document}